\documentclass[structabstract, letter]{aa} 
\usepackage{txfonts}
\usepackage{natbib}
\usepackage{graphicx}
\usepackage{subfigure}
\bibpunct{(}{)}{;}{a}{}{,} 

\begin{document}
\title{Terahertz hot electron bolometer waveguide mixers for GREAT}

\author{P. P{\"u}tz \and C. E. Honingh  \and K. Jacobs \and M. Justen  \and M. Schultz \and J. Stutzki}

\institute{
K\"olner Observatorium f\"ur Submm Astronomie (KOSMA), I. Physikalisches Institut, Universit\"at zu K\"oln, Z\"ulpicher Strasse~77, 50937 K\"oln, Germany \\
\email{puetz@ph1.uni-koeln.de, lastname@ph1.uni-koeln.de}
}

\abstract{
Supplementing the publications based on the first-light observations with the \underline{G}erman \underline{RE}ceiver for \underline{A}stronomy at \underline{T}erahertz frequencies (GREAT) on SOFIA, we present background information on the underlying heterodyne detector technology. This letter complements the GREAT instrument letter and focuses on the mixers itself.} {We describe the superconducting hot electron bolometer (HEB) detectors that are used as frequency mixers in the L1 (1400~GHz), L2 (1900~GHz), and M (2500~GHz) channels of GREAT. Measured performance of the detectors is presented and background information on their operation in GREAT is given.
}
{Our mixer units are waveguide-based and couple to free-space radiation via a feedhorn antenna. The HEB mixers are designed, fabricated, characterized, and flight-qualified in-house. We are able to use the full intermediate frequency bandwidth of the mixers using silicon-germanium multi-octave cryogenic low-noise amplifiers with very low input return loss.}
{
Superconducting HEB mixers have proven to be practical and sensitive detectors for high-resolution THz frequency spectroscopy on SOFIA. We show that our niobium-titanium-nitride (NbTiN) material HEBs on silicon nitride (SiN) membrane substrates have an intermediate frequency (IF) noise roll-off frequency above 2.8~GHz, which does not limit the current receiver IF bandwidth. Our mixer technology development efforts culminate in the first successful operation of a waveguide-based HEB mixer at 2.5~THz and deployment for radioastronomy. A significant contribution to the success of GREAT is made by technological development, thorough characterization and performance optimization of the mixer and its IF interface for receiver operation on SOFIA. In particular, the development of an optimized mixer IF interface contributes to the low passband ripple and excellent stability, which GREAT demonstrated during its initial successful astronomical observation runs.
}
{}

\keywords{instrumentation: detectors -- techniques: spectroscopic -- telescopes: SOFIA}

\maketitle

\section{Introduction}

For very high-resolution far-infrared (FIR) spectroscopy on the Stratospheric Observatory for Infrared Astronomy (SOFIA) the heterodyne observation technique, routinely used at lower frequencies, is extended to THz frequencies. Superconducting hot electron bolometers (HEBs) are used as the frequency mixer because presently they are the only detectors with sufficient sensitivity in the FIR and have a local oscillator (LO) power requirement that is sufficiently low to be met with current solid-state multiplier chains that can be flown on SOFIA. The HEB mixers have been already successfully used in radio astronomy for ground-based observations at the SMT \citep{ Kawamura:2000}, the Smithsonian Center for Astrophysics Receiver Lab Telescope \citep{Meledin:2004}, at APEX \citep{Wiedner:2006, Meledin:2009}, and in space on the Herschel Space Observatory instrument HIFI \citep{deGraauw:2010}.

This paper describes the mixers that are currently used in GREAT\footnote{GREAT is a development by the MPI f\"ur Radioastronomie (Principal Investigator: R. G\"usten) and KOSMA, I.~Physikalisches Institut der Universit\"at zu K\"oln, in cooperation with the MPI f\"ur Sonnensystemforschung and the DLR Institut f\"ur Planetenforschung.}, whose system performance aspects are described in \citet{Heyminck:2012}. We first briefly describe the functionality of a HEB device and of the THz waveguide technology used within the detector units. We continue with a discussion of the cross-calibration uncertainties inherent to the measurement setups in the laboratory and compare our laboratory-measured mixer performance with the measured GREAT performance described in \citet{Heyminck:2012}. Then we describe our measures taken for intermediate frequency (IF) passband optimization and the necessity for direct-detection correction of the mixer's response. Finally, we close with an outlook describing future developments to the GREAT mixers.

\section{Waveguide HEB mixers for GREAT}

The HEB mixers for GREAT are based on a long heritage in the development of waveguide mixers at the University of Cologne and implementation as observatory instrumentation.  We have pushed waveguide technology, which is a standard approach at mm and submm wavelengths and delivers best possible performance, into the THz regime, enabling superior feedhorn optical coupling to the telescope. More common lens-based mixer concepts are easier to implement, but we aimed for advantages that a made possible because of our past experience with these mixer concepts at lower frequencies. We describe these advantages in Sect.~\ref{wg} in more detail.

A first HEB THz waveguide mixer from Cologne at 1.4~THz was in operation in an astronomical receiver at APEX in 2005 \citep{Wiedner:2006} and the technology used for the L1-channel mixer unit on GREAT is very similar to this one. A detailed description of this mixer development can be found in \citet{Munoz:2006}.

Initially the 1.9~THz L2-channel mixers were similar to their L1 counterpart with, essentially, an radio frequency (RF) design scaled to the higher operation frequency being implemented. Owing to the delays GREAT experienced through the SOFIA project, we were able to implement our next-generation, greatly enhanced device design, which is based on beam-lead technology and is described in the Sect.~\ref{HEB}. The main driver for beam-lead technology is that it opens a path to even higher operation frequency waveguide devices and also brings advantages for future focal plane array (multi-pixel) heterodyne instruments, such as upGREAT (Sect.~\ref{outlook}).

The 2.5~THz M-channel mixer is a direct consequence of this beam-lead technology. It is the first waveguide-based HEB mixer operating at 2.5~THz published that has been successfully employed in a receiver producing astrophysical data. The underlying design and fabrication of the beam-lead mixers is published in \citet{Puetz:2011} and a much more detailed description of the 2.5~THz mixers will be published in \citet[in prep.]{Puetz:2012}.

\subsection{HEB devices}
\label{HEB}

\begin{figure}
 \resizebox{\hsize}{!}{\includegraphics{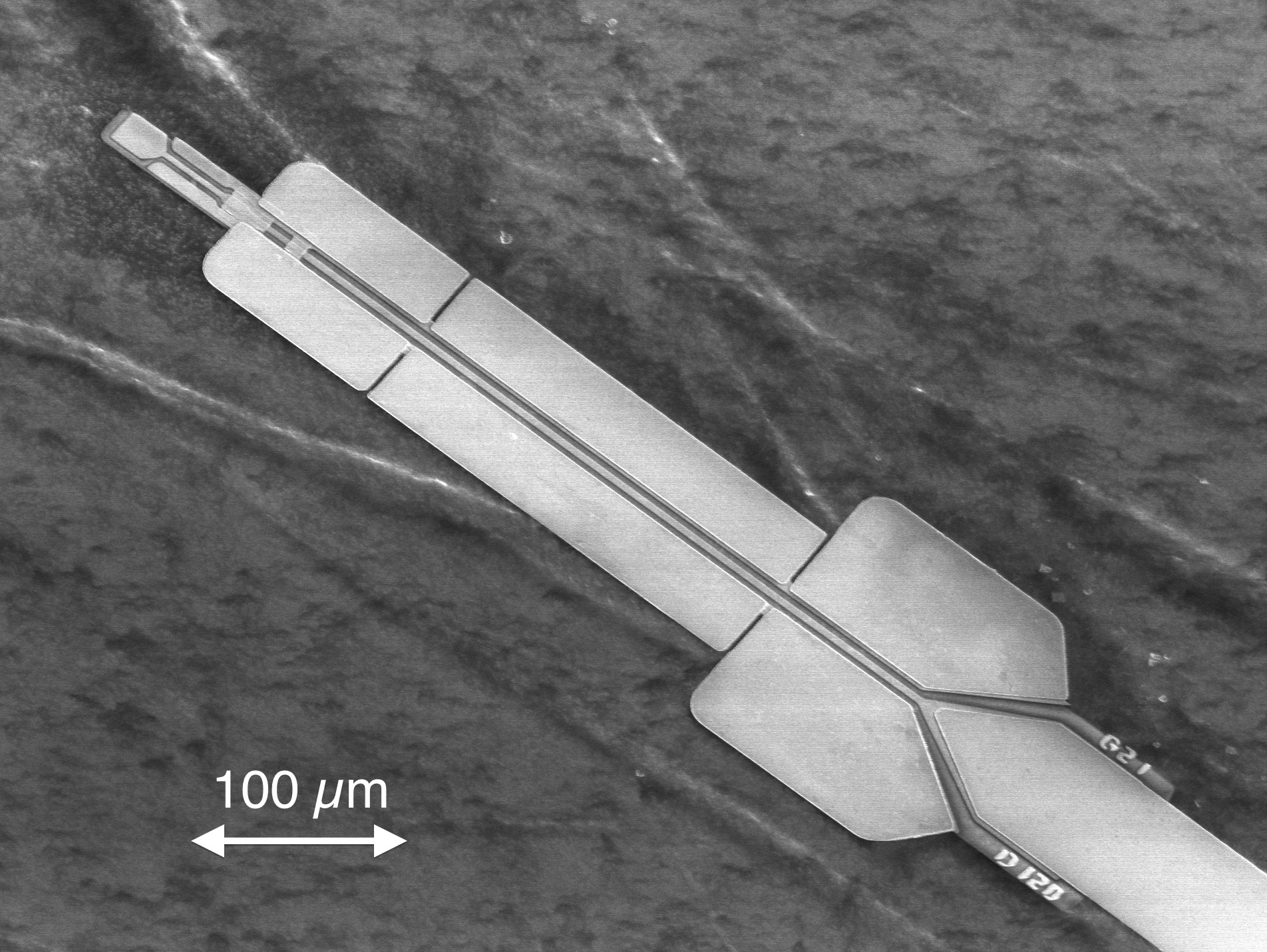}}
  \caption{HEB device: SEM micrograph of a 2.5~THz HEB device with SiN substrate and beam leads (large light-colored features). The scale bar corresponds to 100~$\mu$m.}
  \label{sembildheb}
\end{figure}
A phonon-cooled hot electron bolometer (HEB) device consists of an ultra thin (few nm) microbridge of superconducting material. This strip can function as a bolometer if it is biased at its superconducting transition temperature (T$_\mathrm{c}$), where small temperature variations lead to a strong change in resistance \citep{Goltsman:1995}. The main cooling path of the bolometer is via the electron-phonon interaction in the microbridge and subsequent escape of the phonons into the substrate. The time constant of this interaction, depending on the superconducting material, the substrate material and the interfaces both to the substrate and the electrical contacts \citep{Baselmans:2005} can be such that the bolometer follows temperature variations with a frequency well into the GHz range. Used as a square-law mixer for THz radiation, the beat note between a local oscillator source and the THz signal can thus be used to down-convert the THz signal to a IF in the few GHz range. A crucial advantage of superconducting HEB mixers is the small amount of local oscillator power required in the order of a few hundred nW. The LO power consumption is a function of the volume of the bridge, which can be adapted during fabrication to a certain extent.

The instantaneous IF bandwidth (3~dB roll-off) of the HEB mixer is mainly determined by the time constant of the electron-phonon interaction time of the bolometer. Therefore it is important that a well-matched bridge material-substrate material combination is chosen and that the fabrication parameters of the thin superconducting detector layer are optimized for a maximum IF roll-off frequency. For our waveguide mixers the circuit design makes using very thin (a few micrometers) substrates mandatory. We used 2~$\mu$m low-stress silicon nitride (SiN) membrane layers deposited on a 500~$\mu$m thickness silicon handle wafer for easier processing. During the final stages of fabrication the handle Si was back-etched and the thin SiN layer cut through deep reactive-ion processing. For the bolometer microbridge layer material we used niobium-titanium-nitride (NbTiN) with a thickness of 4--5~nm, a sheet resistance R$_\textrm{s} = 300~\Omega / \square$ and with \textbf{a} critical temperature T$_\textrm{c} = 8.5$~K for the initial NbTiN layer and about 8~K for the completed device.

Owing to the small lateral dimensions with a typical length of 400~nm and widths that vary between 1--6~$\mu$m, the HEB microbridge needs to be defined through e-beam lithography. The width is set by the normal impedance level required for the specific embedding RF circuit and also the available LO power. M-channel HEB devices are purposely smaller in volume because of the LO power limitation. The HEB microbridges for the GREAT L1-channel are typically 6.2~$\mu$m wide, corresponding to a normal state resistance R$_\textrm{N} = 20~\Omega$, for the L2 channel 3.1~$\mu$m wide with R$_\textrm{N} = 40~\Omega$, and 1.1 or 1.6~$\mu$m wide, corresponding to R$_\textrm{N} = 120$ or 80~$\Omega$ for the M channel. 

The HEB microbridge is embedded into a planar on-chip RF circuit, which provides the waveguide probe antenna, transmission lines, a matching circuit and the IF output. To improve the device performance and also cryo-mechanical cycling durability, we now integrated beam-lead structures for device mounting and RF, IF, and DC electrical contacts. The SEM micrograph in Fig.~\ref{sembildheb} shows a 2.5~THz device. The beam leads are the large light-colored features with the device and waveguide probe antenna at the top left. The IF and DC beam lead at the bottom right extends well beyond the micrograph with a total length of 1800~$\mu$m.  All device fabrication including the NbTiN thin film is made in-house.

\subsection{THz waveguide mixers}
\label{wg}

\begin{figure}
 \resizebox{0.495 \hsize}{!}{\includegraphics{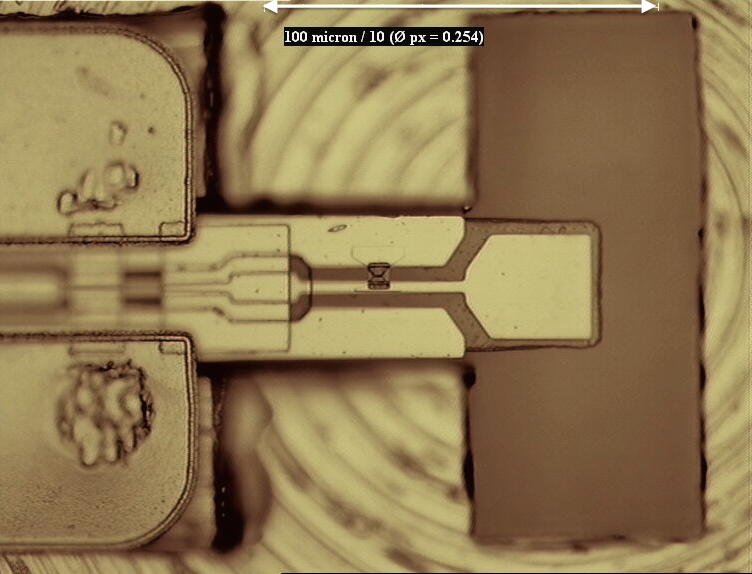}} \resizebox{0.483 \hsize}{!} {\includegraphics{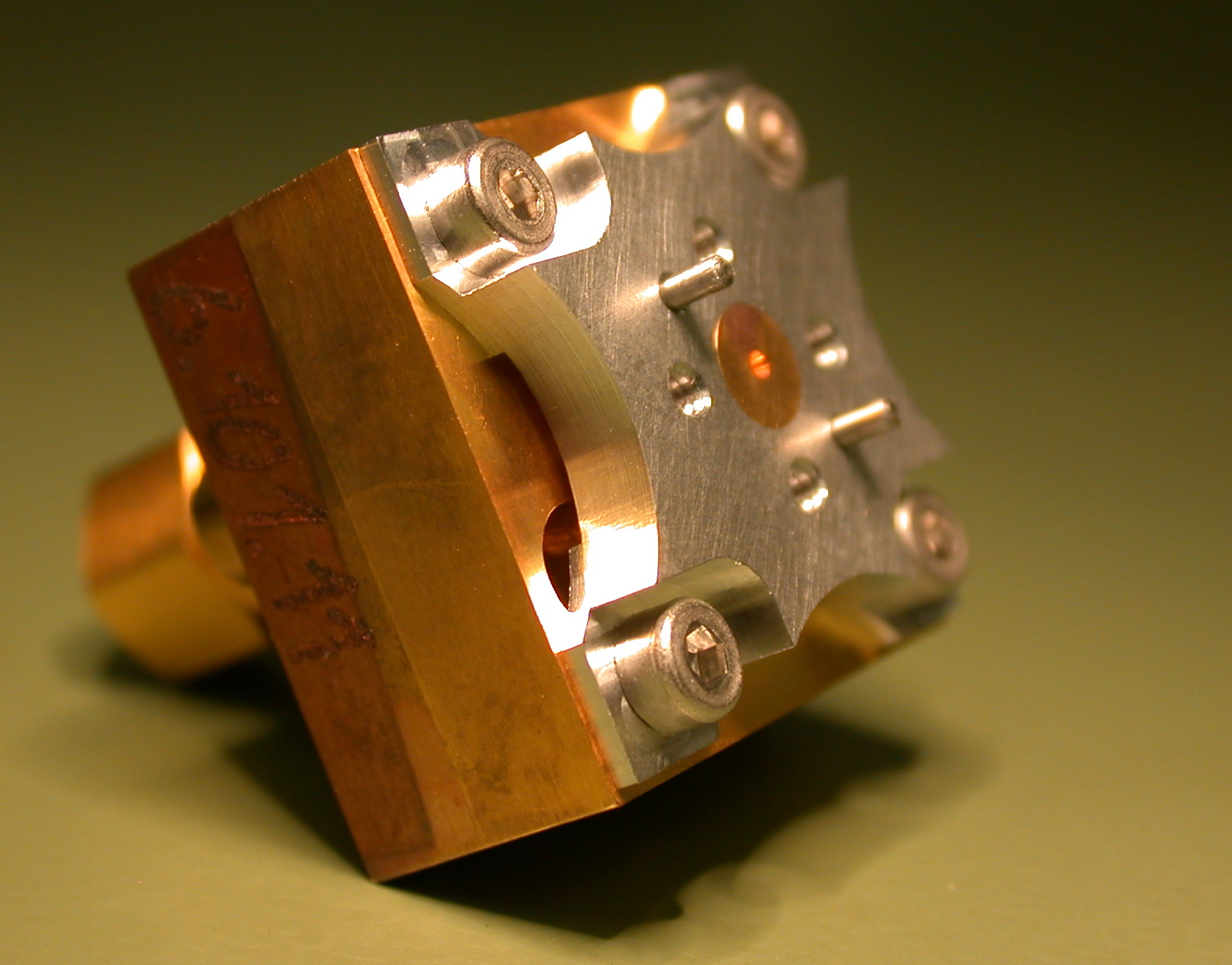}}
  \caption{Waveguide Mixer: (Left) Optical microscope image of a 1.9~THz device mounted into a waveguide backshort block. The scale bar corresponds to 100~$\mu$m. (Right) Flight hardware mixer unit with components: metal backshort block, IF connector and horn clamp with press-fit feedhorn and dowel pins for precise alignment to the cold receiver optics. The foot print of the mixer unit is (20~mm)$^2$.}
  \label{mixer}
\end{figure}

Because they are much smaller than the wavelength, HEB devices have to be connected to an antenna to couple to free space radiation. Planar antennas integrated with the HEB device are the traditional approach to design and fabricate HEB mixers. They use thick Si substrates on hyper-hemispherical lenses, which can easily be handled and mounted. These mixers show excellent sensitivity, see for example \citet{Cherednichenko:2008}, but the Gaussian beam parameters are not easily matched to external optical components required for coupling to a telescope. In particular, the alignment of the mixer beam with respect to the receiver optical axis is not trivial and can result in beam squint.

The alternative is embedding the HEB into a waveguide mixer block using a very thin dielectric substrate with an integrated waveguide probe antenna and a waveguide feedhorn for the coupling to the free space. This decouples the optical alignment from the positioning of the tiny detector chip. Mixers can then be exchanged quickly without time-consuming receiver optics re-alignment with obvious advantages for implementation of a focal plane array of mixers.

The practical realization of THz waveguide mixers is challenging, not only because of the small dimensions of the wave\-guide (96~$\mu$m x 48~$\mu$m at 2.5~THz) and substrate channel, but also because the HEB device substrates need to be very thin (few micrometers) to avoid THz waves propagating in the substrate channel. KOSMA has acquired substantial know-how in the machining of these tiny metal waveguide features using direct-milling, stamping and cutting techniques. As a feedhorn we use a proprietary spline-profile with excellent beam characteristics \citep{Rabanus:2006}, which was fabricated at RPG\footnote{Radiometer Physics GmbH, http://www.radiometer-physics.de}.

The beam leads serve as registration elements with respect to the metal features of the block, which are machined to a few micrometers accuracy. The waveguide probe antenna of the device has to be positioned to $\pm 2~\mu$m accuracy for optimum performance of the 2.5~THz device, achieved via a micro-manipulator setup.

\section{Performance measurements}

Each HEB mixer is characterized and qualified for GREAT through measurements in a laboratory setup using a small test cryostat. The setup is optimized for a simplified optical path to reduce calibration uncertainties arising from the optical and atmospheric transmission losses (mainly at 1.9~THz and 2.5~THz), and therewith helps to deduce the intrinsic mixer noise performance.

For the L-channel mixer measurements in Cologne we can evacuate most parts of the optical path outside of the test cryostat including the Martin-Puplett polarizing diplexer, which is necessitated by the low output powers of the synthesizer-driven multiplier based LO chains. In principle this is similar to what is used in the GREAT receiver. The measured noise temperatures in the laboratory and in GREAT for the same mixer generally agree well.

With the availability of a high output power LO such as a gas laser, the diplexer can be replaced by a thin dielectric foil beamsplitter. In combination with a completely evacuated optical path including the calibration loads, the contribution of the setup to the measured noise temperature is not significant and evaluation of the mixer noise performance is much more straightforward. A setup like this was used for the M-channel mixer characterization.

\subsection{Sensitivity}

 \begin{figure}
\resizebox{\hsize}{!}{\includegraphics{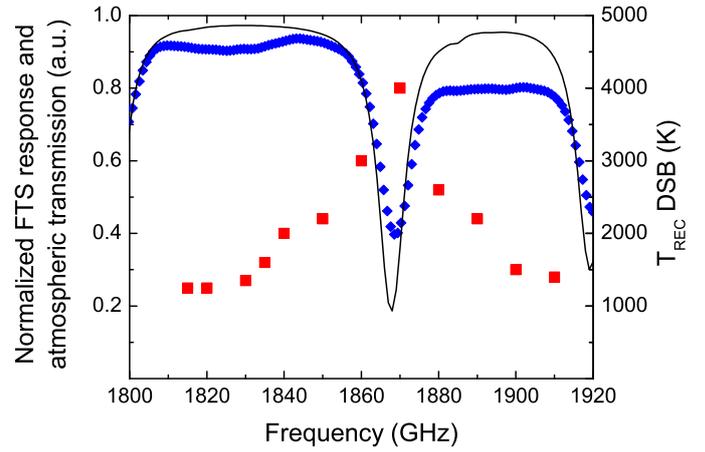}}
  \caption{Measured spectral response of the L2-channel flight mixer: The red squares are the uncorrected T$_\textrm{rec}$\,(DSB) vs. local oscillator frequency measured with the GREAT system at the ground-support facility in Dryden. The blue diamonds are the normalized direct-detection response measured in a test cryostat with the FTS setup at the Cologne lab. The absorption feature in the FTS response at 1868~GHz is caused by water in the beam path. The black solid line is the calculated atmospheric transmission for the FTS setup in Cologne.
}
  \label{FTS}
\end{figure}

 \begin{figure}
\resizebox{\hsize}{!}{\includegraphics{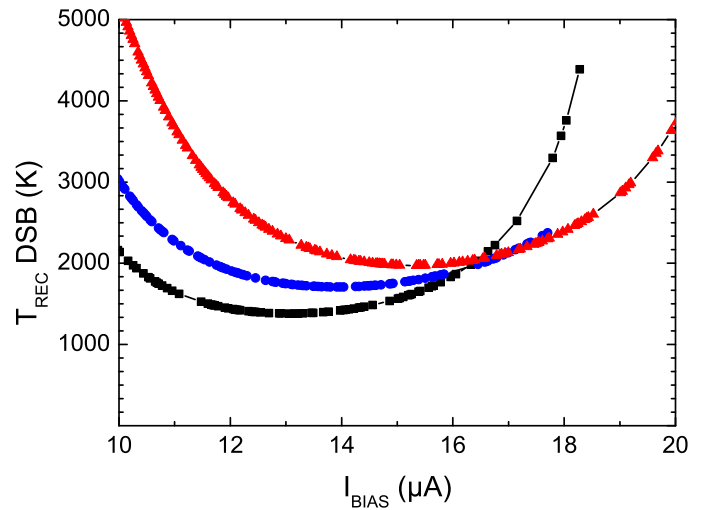}}
  \caption{Series of measured T$_\textrm{rec}$\,(DSB) vs. mixer bias current $\mbox{I}_\textrm{bias}$ curves at three different bias voltages for the 2.5~THz mixer used in the M-channel receiver during the 2011 operations. The mixer was measured in a laboratory test setup using a FIR gas laser at 2523~GHz. From top to bottom the traces are for bias voltages 0.65~mV, 0.5~mV and 0.4~mV. For these measurements the pump power of the FIR gas laser LO was varied by means of an attenuating wire grid.
}
  \label{120_ohm}
\end{figure}

Measured T$_\textrm{rec}$ in the laboratory and GREAT agree well for the lowest frequency L1 channel. In GREAT T$_\textrm{rec} = 1000-1750$~K\,(DSB) are measured across the bands defined by the L1$_\textrm{a}$ and L1$_\textrm{b}$ configurations \citep{Heyminck:2012}. More details on the underlying technology of the L1 mixer are found in \citet{Munoz:2006}.

At 1.9~THz the broadband Fourier Transform Spectroscopy (FTS) response of the L2-channel flight mixer in Fig.~Ê\ref{FTS} shows that the sensitivity variations of the mixer are small over the tuning range of the LO of the L2 channel. In the FTS response the influence of water vapor in the 5~cm long optical path in ambient air is evident and an estimated transmission along this path is plotted.\footnote{$am$, CfA, https://www.cfa.harvard.edu/~spaine/am/index.html} The measurements of T$_\textrm{rec}$ with GREAT show similar variations that cannot be caused by a change in mixer sensitivity alone. With GREAT minimum T$_\textrm{rec} = 1500$~K are measured for LO frequencies around 1900~GHz near the important [C~II] transition fully consistent with the laboratory characterization measurements and thus confirming optimum performance of the GREAT receiver internal optics.

The measurement of the M-channel 2.5~THz mixer was performed within an EU\,FP7 collaboration\footnote{Radionet, http://www.radionet-eu.org/} using a FIR gas laser as local oscillator at 2.523~THz at SRON Groningen, the Netherlands. A 2~$\mu$m mylar beamsplitter was used as diplexer, and the optical path between the mixer and the calibration loads was evacuated with no vacuum window in-between. This setup yielded a direct-detection compensated (see Sect.~\ref{dd}) best receiver noise temperature T$_\textrm{rec} = 1400$~K (DSB)  at an IF of 1.25~GHz. Fig.~\ref{120_ohm} shows the measured T$_\textrm{rec}$ vs. I$_\textrm{bias}$ for three different mixer bias voltages with the output (pump) power of the LO being varied by means of an attenuating wire grid. It can be seen that this mixer demonstrates a noticeable mixer bias voltage, V$_\textrm{bias}$, dependency of T$_\textrm{rec}$ and for V$_\textrm{bias} = 0.65$~mV as used in the GREAT M channel during the science flights a minimum noise temperature T$_\textrm{rec} = 2000$~K (DSB) is measured. 

In GREAT more elaborate optics are required for coupling of the signal and LO. In particular, the synthesizer-driven solid-state LOs used on GREAT require a Martin-Puplett polarizing diplexer for efficient power coupling. For the M channel the total optical throughput in front of the mixer is estimated to 70\,\% \citep[priv. comm.]{Heyminck2:2012}. The measured single-sideband T$_\textrm{rec} = 7000$~K  (corresponding to 3500~K DSB) for the M$_\textrm{a}$ configuration at 2509 GHz is consistent with the laboratory measurements due to the non-identical receiver configurations \citep{Heyminck:2012}. Mainly the optics losses themselves contribute an additional 1500~K to the receiver noise, increasing T$_\textrm{rec}$ to approximately 3500~K (DSB). A more elaborate paper on the KOSMA 2.5~THz waveguide mixer development and laboratory characterization will be published in \citet[in prep.]{Puetz:2012}.

\subsection{Instantaneous IF bandwidth and stability}

\begin{figure}
\resizebox{\hsize}{!}{\includegraphics{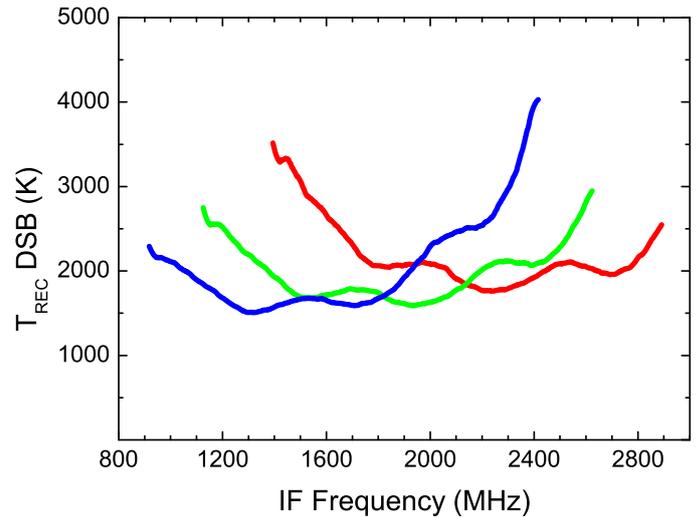}}
  \caption{Uncorrected T$_\textrm{rec}$\,(DSB) vs. intermediate frequency (IF) of the L2-channel at a LO frequency of 1821~GHz measured during GREAT system tests at the MPIfR lab: The blue, green and red lines are the measured T$_\textrm{rec}$ (DSB) for the Martin-Puplett diplexer set at half path-length differences 45, 35 and 30~mm, respectively. The T$_\textrm{rec}$ is dominated by the diplexer passband and a 3~dB noise roll-off of the HEB mixer $> $ 2.8~GHz is measured.}
  \label{IF_rolloff}
\end{figure}

The instantaneous IF bandwidth of a phonon-cooled HEB mixer is determined by the microbridge material, its interface to the substrate as well as the normal metal contact pads and is limited to a few GHz \citep{Baselmans:2005}.

To fully exploit the limited IF bandwidth of HEB mixers, the cryogenic low-noise amplifier (LNA) following the mixer needs to have a multi-octave bandwidth. Conventional LNA show a poor input match under these conditions, causing standing waves between mixer and amplifier that lead to baseline ripples and instability in the spectra.  
It would therefore be convenient and prudent to insert a high-frequency isolator between the mixer and the LNA to suppress the standing waves. 
Unfortunately, commercially available cryogenic isolators have less than one octave bandwidth and thus limit the available bandwidth unacceptably in the HEB IF range \citep{Wiedner:2006}. For GREAT, we fortunately profited from the development of cryogenic wideband LNAs using silicon germanium transistor technology that was developed with the THz HEB mixer application in mind \citep{Weinreb:2009}.  The input return loss of these amplifiers at 15~K operating temperature is measured to be below -12~dB  from 0.5--5.0~GHz with an excellent input noise temperature \citep{Bardin:2009}. The return loss of the mixer is also designed to be equally low, so that the standing wave between mixer and LNA is minimized. Laboratory measurements of the IF passband confirmed the increased usable IF bandwidth for GREAT \citep{Puetz:2011}.

During the final system end-to-end tests with GREAT it was found that even with this improved IF match residual baseline ripples of several periods appeared on the spectra during long integrations and measurably affected the system's baseline quality and therewith the quality of and confidence in the reduced data. The ripple period corresponded to the 10~cm IF coaxial cable length between mixer unit and LNA. The ripple is presumably generated by small impedance changes at the mixer IF output port caused by minute receiver system instabilities that influence the mixer bias \citep{Kooi:2007}. To increase the ripple period in the IF we even more minimized the connection length between the mixer unit and the LNA for each channel of GREAT. Using a newly developed external bias-T unit the total electrical length now is only 2~cm. The result is that the residual ripple amplitude now shows only half a period in the spectrometer band, which greatly facilitates the baseline correction of astronomical data, especially for wide spectral line features \citep{Heyminck:2012}.

For our devices using a NbTiN microbridge and a 2~$\mu$m thin SiN membrane as substrate we can confirm that the 3~dB IF noise roll-off frequency is sufficient for the present L1 and L2 channels. For the measurements in Fig.~\ref{IF_rolloff} the IF bandpass of the Martin-Puplett diplexer, which evidently is smaller than the instantaneous IF bandwidth of the mixer, was adjusted to the best IF center frequency for each scan. Apart from a long wavelength ripple in the IF passband the noise roll-off frequency of larger than 2.8~GHz is visible.

\subsubsection{Direct-detection}
\label{dd}

\begin{figure}
\resizebox{\hsize}{!}{\includegraphics{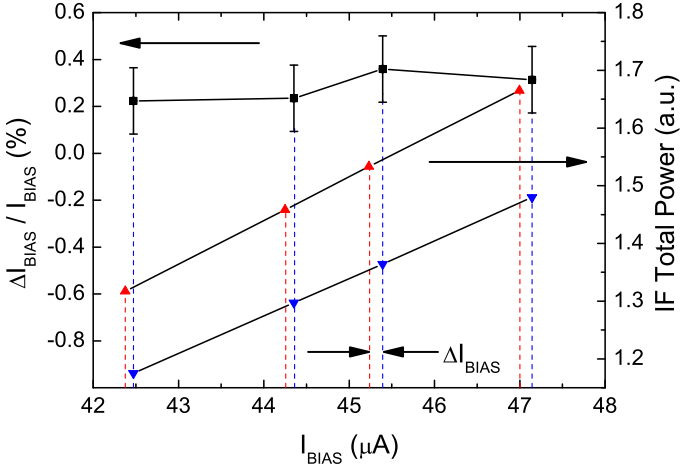}}
  \caption{L2-channel IF power (right axis) for hot (upward pointing triangles) and cold (downward pointing triangles) load measurements taken at four different mixer bias current settings, showing the mixer gain dependence on the mixer bias current and the mixer current change (dashed lines) due to the direct detection of the broadband thermal radiation of the loads. The relative current change (squares) is given on the left axis. The bias voltage is constant throughout the experiment.}
  \label{dd_effect}
\end{figure}

A HEB mixer also exhibits a direct-detection (bolometric) response to changes in thermal radiation, e.g. from switching between hot and cold calibration loads. It is important to understand the effects of this direct-detection on the line temperature calibration as well as on the determination of the receiver sensitivity, which is required for calculation of on-source integration times. As an example, Fig.~\ref{dd_effect} shows the effect for the L2 channel where the IF output power values measured on the hot load (upward pointing triangles) and cold load (downward pointing triangles) are plotted for four mixer bias current values at constant optimum bias voltage. The mixer bias current is set to a fixed value (by varying the local oscillator power) at each of the four cold load measurements. Clearly, the mixer gain depends almost linear on the mixer bias current within the limits of the plot. 

When switching from the cold load to the hot load at a given current value (at constant local oscillator power), the IF power obviously changes due to the temperature of the load. This is the heterodyne response of the mixer. At the same time, the mixer bias current changes slightly ($\Delta \mbox{I}_\textrm{bias}$) between the hot load measurement and the cold load measurement, as indicated by the vertical dashed lines. This is the direct-detection effect caused by the thermal broadband radiation that heats the bolometer. The relative mixer bias current change $\Delta \mbox{I}_\textrm{bias} / \mbox{I}$ due to this direct-detection effect is plotted on the left scale. The error bar shows the accuracy of the mixer current measurement in the GREAT instrument.

The small difference in mixer bias current (about 0.2\,\%) for the hot and cold measurement results in a mixer gain difference for the two measurements, in turn resulting in a deviation from the linear relation between the temperature of the loads and the IF output power that is usually assumed for heterodyne receivers. From the plot, the relative gain change due to the hot/cold mixer current difference can be determined (in the 0.5\,\% range for the shown L2-channel) for a given mixer bias current and used in an analysis of the nonlinearity, so that its effect on line temperature calibration can be eliminated. An analysis covering the basic effect can be found in \citet{Baselmans:2006}, \citet{Lobanov:2009}, and \citet{Cherednichenko:2007}, and a more complete analysis taking the different noise regimes of HEB mixers into account and tailored to astronomical applications of HEB receivers is beyond the scope of this paper and is in preparation for separate publication \citep{Jacobs:2012}.

The magnitude of the effect essentially scales with the volume of the HEB device. The device volumes for the GREAT channels vary as $\mbox{L1} > \mbox{L2} > \mbox{M}$. As a result, the GREAT L1-channel shows a negligible direct detection effect. For the L2-channel, our analysis results in a line temperature calibration correction of about 1.5\,\% at the bias conditions used in the science flights. For the M-channel the effect is stronger (5--9\,\%). For the line calibration, the effect of the nonlinearity is relatively small because the calibration depends on the difference of the hot and cold IF output powers with the calibration factor being $\propto (\mbox{T}_\textrm{hot} - \mbox{T}_\textrm{cold}) / (\mbox{P}_\textrm{hot} - \mbox{P}_\textrm{cold})$. The receiver noise temperature, determined by the y-factor method $\mbox{T}_\textrm{rec} = (\mbox{T}_\textrm{hot} - \mbox{y} \cdot \mbox{T}_\textrm{cold}) / (\mbox{y} - 1)$, where $\mbox{y} = \mbox{P}_\textrm{hot} / \mbox{P}_\textrm{cold}$, is much more affected by the gain change between hot and cold measurements because it depends on the ratio of the hot and cold IF power values. The effect is especially strong for noise temperatures in the present few 1000\,K range where y approaches close to unity. For example, the noise temperature of the M channel would be overestimated by up to 50\,\% if the y-factor would not be corrected for the direct-detection.

For the GREAT basic science flights, the direct-detection effect was eliminated in the instrument by adjusting the local oscillator power levels at the hot, cold and sky calibration measurements such that the mixer current (and mixer gain) remains constant \citep{Heyminck:2012}. The spectra therefore do not need any additional corrections in their temperature calibration.

\section{Outlook}
\label{outlook}

Setting the receiver diplexer bandpass limitation aside, the instantaneous IF bandwidth of the present mixers is still marginal for galactic center or extragalactic THz sources, and the situation worsens toward higher operation frequencies, such as the upcoming H-channel on GREAT at 4.7~THz, the astronomically important [O I] line. For HEB mixers based on niobium nitride (NbN) instead of NbTiN, wider bandwidths have been reported, but only on Si substrates. Using silicon on insulator (SOI) techniques, we now have the technology to produce the very thin Si membrane substrates needed for NbN HEB waveguide devices, setting the path to wider IF bandwidths in the future. In addition, we will further investigate waveguide machining technology based on Si micro machining using deep-reactive ion-etching. With refined device assembly techniques we have started the development of small THz arrays. For SOFIA the upGREAT array receiver with 2x 7 pixels (1.9--2.5~THz) and 7 pixels (4.7~THz) is funded and under development \citep{Heyminck:2012}.

\begin{acknowledgements}

HEB mixer development is supported by the Deutsches Zentrum f\"ur Luft- und Raumfahrt (DLR) under grant numbers 50\,OK\,801 (GREAT) and 50\,OK\,1103 (upGREAT), and is carried out within the Collaborative Research Council 956, sub-project D3, funded by the Deutsche Forschungsgemeinschaft (DFG). The 2.5~THz mixer measurement at SRON, Groningen, was funded through EU\,FP7, Radionet, Advanced Radio Astronomy in Europe, grant 227290.

We thank Oliver Ricken (KOSMA) for his important contribution during GREAT receiver testing and the measured flight mixer data. We thank our machine shop for their dedication and excellence in machining our THz mixer blocks. We thank J.R. Gao (SRON Groningen and Kavli Institute of Nanoscience, TU Delft), W. Zhang and W. Miao (both Purple Mountain Observatory, China) and M. Brasse (KOSMA) for their assistance with the measurements at the 2.5~THz laser setup at SRON Groningen. 

\end{acknowledgements}

\bibliographystyle{aa}
\bibliography{THz_HEB_mixer_GREAT}

\end{document}